\documentclass[prl,twocolumn,showpacs]{revtex4}
\usepackage{amsmath,amsgen,amstext,amsbsy,amsopn,amsthm,amssymb,epsf}

\def\be{\begin{equation}}
\def\ee{\end{equation}}
\def\ba{\begin{align}}
\def\bm{\begin{multline}}
\def\bfig{\begin{figure}[htb]}
\def\efig{\end{figure}}

\DeclareMathSymbol{\leqslant}{\mathalpha}{AMSa}{"36}
\DeclareMathSymbol{\geqslant}{\mathalpha}{AMSa}{"3E}
\DeclareMathSymbol{\doteqdot}{\mathalpha}{AMSa}{"2B}
\DeclareMathSymbol{\circlearrowright}{\mathalpha}{AMSa}{"08}
\DeclareMathSymbol{\subsetneq}{\mathalpha}{AMSb}{"28}
\DeclareMathSymbol{\supsetneq}{\mathalpha}{AMSb}{"29}
\renewcommand{\leq}{\;\leqslant\;}
\renewcommand{\geq}{\;\geqslant\;}

\newcommand{\dd}{{\rm d}}
\newcommand{\e}[1]{\,{\rm e}^{#1}\,}

\newcommand{\sumtwo}[2]{\sum_{\substack{#1 \\ #2}}}

\newcommand{\upchi}{\raise 2pt \hbox{$\chi$}}

\def\writefig#1 #2 #3 {\rlap{\kern #1 truecm \raise #2 truecm
\hbox{#3}}}
\def\figtext#1{\smash{\hbox{#1}} \vspace{-5mm}}

\newcommand{\bsc}{{\boldsymbol c}}

\newcommand{\bsrho}{{\boldsymbol\rho}}
\newcommand{\bsvarrho}{{\boldsymbol\varrho}}
\newcommand{\bssigma}{{\boldsymbol\sigma}}

\begin{document}

\title{The relation between Feynman cycles and off-diagonal long-range order}

\author{Daniel Ueltschi}
\affiliation{Department of Mathematics,
University of Arizona,
Tucson, AZ 85721, USA}

\date{August 23, 2006}

\begin{abstract}
The usual order parameter for the Bose-Einstein condensation involves the off-diagonal correlation function of Penrose and Onsager, but an alternative is Feynman's notion of infinite cycles. We present a formula that relates both order parameters. We discuss its validity with the help of rigorous results and heuristic arguments. The conclusion is that infinite cycles do not always represent the Bose condensate.
\end{abstract}


\maketitle

\subsection{Introduction}

In 1953 Feynman suggested an order parameter for the Bose-Einstein condensation of interacting systems \cite{Fey}. It involves the lengths of space-time trajectories in the Feynman-Kac representation. Three years later Penrose and Onsager introduced the concept of off-diagonal
long-range order \cite{PO}. The latter is now accepted as the correct order parameter. Yet
Feynman cycles remain actual because they are easier to deal with, and because they
provide a useful microscopic heuristic \cite{Cep,GCL,Sch}. Another order parameter that
involves space-time trajectories winding in spatial directions --- rather than in the
imaginary time direction --- is closely related to Feynman cycles; it has been used
extensively in investigations of the supersolid phase, see \cite{BSZK,SPATS} and
references therein.

The theory of the ideal gas has led to the expression ``Bose condensate'' to designate the
particles that occupy the zero Fourier mode. The concept extends to interacting systems
via the off-diagonal correlation function $\bssigma(x,y)$. The density of the Bose
condensate is then
$$
\bsrho_0 = \lim_{V\to\infty} \Big\langle \frac{N_0}V \Big\rangle =
\lim_{|x-y|\to\infty} \bssigma(x,y).
$$
Feynman's approach introduces the density of infinite cycles, $\bsvarrho(\infty)$, to be precisely defined below. The natural question is whether these densities are identical.

The links between Feynman cycles and off-diagonal long-range order were explored by S\"ut\H o, who proved that infinite cycles occur in the ideal gas below the
critical temperature \cite{Suto}, and that they do not occur above it \cite{Suto2}. These
results were later extended to the mean-field Bose gas \cite{BCMP,DMP}. Articles
\cite{Cep,GCL,Sch,BSZK,SPATS,Suto,Suto2,BCMP,DMP} all assume, or conjecture, that $\bsrho_0 = \bsvarrho(\infty)$ in any Bose system with reasonable interactions. We will see, however, that this cannot be true in general.

The purpose of this letter is to present an explicit relation between the off-diagonal correlation function and the densities of Feynman cycles. Namely,
\be
\label{LaFormule}
\bssigma(x,y) = \sum_{n\geq1} \bsc_n(x-y) \, \bsvarrho(n) + \bsc_\infty(x-y) \, \bsvarrho(\infty).
\ee
Here, $\bsc_n(x)$ are coefficients, and $\bsvarrho(n)$ denotes
the density of particles in cycles of length $n$. We expect that
$$
0 \leq \bsc_n(x) \leq 1, \quad \text{and} \quad \lim_{n\to\infty} \bsc_n(x) =
\bsc_\infty(x),
$$
for any fixed $x$. The latter statement is not obvious. Notice that
the analogous statement for $\bsvarrho(n)$ is false in general: We
always have $\lim_n \bsvarrho(n) = 0$, but it is possible that
$\bsvarrho(\infty) \neq 0$. In addition, we should have \be
\label{limncoeff} \lim_{|x|\to\infty} \bsc_n(x) = 0, \ee for any
fixed $n$. But $\bsc_\infty(x)$ may converge to a strictly positive
constant $\bsc$. The formula and these properties can be established
in the case of the ideal gas in the canonical ensemble \cite{Uel}.
It is found that
$$
\bsc_n(x) = \e{-x^2/4n\beta}, \quad\quad \bsc_\infty(x) = 1,
$$
and
$$
\bsvarrho(n) = \frac{\e{\beta\mu n}}{(4\pi n\beta)^{d/2}}, \quad\quad \bsvarrho(\infty) =
\max(0,\rho-\rho_{\rm c}).
$$
Here, $\mu$ is the chemical potential (that depends on the density
$\rho$), and $\rho_{\rm c}$ is the critical density of the ideal
gas. Notice that $\bsc=1$.

The interacting gas constitutes a formidable challenge to theoretical physicists. The sole
mathematical proof about the occurrence of Bose-Einstein condensation deals with the
hard-core lattice model at the symmetry point \cite{DLS,KLS}. A weakly interacting system is widely expected to display a Bose-Einstein condensation at low temperature. However, the question of whether interactions increase or decrease the critical temperature is still currently debated; the discussion in \cite{GCL} illustrates this point by quoting several references that draw contradictory conclusions. In recent years progress has been made in understanding Bogolubov's theory \cite{ZB}. Remarkable results dealing with the ground state of low density systems are reviewed in \cite{LSSY}.

Two results can be rigorously derived that add credence to the formula \eqref{LaFormule}. First,
for $\mu<0$, for any $\beta$, and for any repulsive interactions, we have
$$
\lim_{|x-y|\to\infty} \bssigma(x,y) = 0, \quad \text{and} \quad \bsvarrho(\infty) = 0.
$$
Second, a rigorous cluster expansion allows to establish Eq.\ \eqref{limncoeff} in the
regime where $\mu<0$, and where $\beta$ is small with respect to the interactions. These results are proved in \cite{Uel}.

In this letter we provide a heuristic discussion of interacting systems, and we will
conclude that
\begin{itemize}
\item $\bsrho_0=\bsvarrho(\infty)$ when interactions are weak;
\item $\bsrho_0=0$, $\bsvarrho(\infty)>0$ in a crystal at sufficiently low
temperature.
\end{itemize}
It is not clear whether a regime of parameters exists where the constant $\bsc$ differs from 0 and 1.

Throughout this letter, we denote {\it finite volume} expressions in {\it plain characters}, and
{\it infinite volume} expressions in {\it bold characters}.

\subsection{Feynman-Kac expressions}

In this section we introduce the relevant mathematical expressions in the Feynman-Kac representation of Bose systems.
We consider a system of identical bosonic particles interacting with a pair potential. We study this system in the grand-canonical ensemble using the Feynman-Kac representation. The present discussion is necessarily brief, and we refer to \cite{Uel} for a more extended mathematical description.

The grand-canonical partition function in a domain $D$ (a
$d$-dimensional cube of size $L$, volume $V=L^d$, with periodic
boundary conditions) can be expressed using the Feynman-Kac
representation \cite{Gin}. With $\beta$ the inverse temperature and
$\mu$ the chemical potential, it is given by
\bm
\label{FKpartfct}
Z
= \sum_{k\geq1} \frac1{k!} \prod_{j=1}^k \biggl[ \sum_{n_j \geq 1}
\frac{\e{\beta\mu
n_j}}{n_j} \int_D \dd x_j \int \dd W^{n_j \beta}_{x_j x_j}(\omega_j) \\
\e{-\beta U(\omega_j)} \biggr] \prod_{1\leq i<j\leq k}
\e{-\beta U(\omega_i,\omega_j)}.
\end{multline}
This expression is illustrated in Fig.\ \ref{figfeykac}. In words, the partition function
involves a sum over the number $k$ of space-time cycles. For each cycle $j$, we sum over
the winding number $n_j$; we integrate over the initial position $x_j$; we integrate over
a Brownian trajectory $\omega_j$ that starts and ends at $x_j$, winding $n_j$ times around
the time direction. Here, $\dd W_{xy}^\beta(\omega)$ is the Wiener measure for a
trajectory $\omega : [0,\beta] \to D$ such that $\omega(0)=x$ and $\omega(\beta)=y$.
Given a trajectory $\omega$ with winding number $n$, we denote by $U(\omega)$ the
interactions between its legs. That is,
$$
U(\omega) = \sum_{0\leq i<j\leq n-1} \frac1\beta \int_0^\beta U \bigl(
\omega(i\beta+s) - \omega(j\beta+s) \bigr) \dd s.
$$
Different trajectories $\omega$ and
$\omega'$ (with winding numbers $n$ and $n'$) have interactions
$$
U(\omega,\omega') = \sumtwo{0\leq i\leq n-1}{0\leq j\leq n'-1} \frac1\beta \int_0^\beta U \bigl(
\omega(i\beta+s) - \omega'(j\beta+s) \bigr) \dd s.
$$
In the two equations above, $U(x)$ denotes the pair interaction potential between two particles
separated by a distance $|x|$. We suppose that $U(x)$ is nonnegative and spherically
symmetric.

\bfig
\epsfxsize=70mm
\centerline{\epsffile{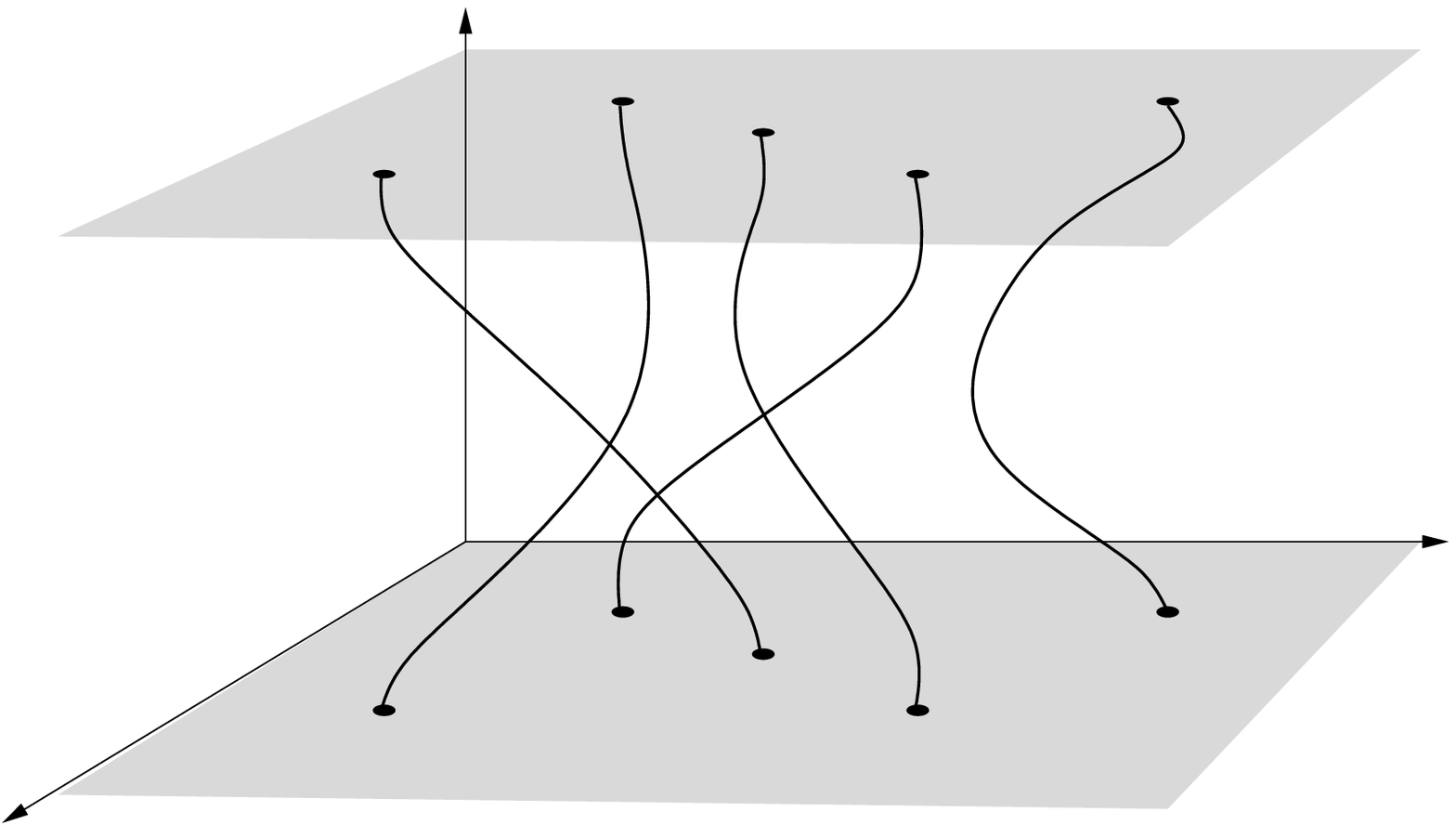}}
\figtext{
\writefig   0.4 0.35    {$x$}
\writefig   7.8 1.65    {$y$}
\writefig   2.55    4.2 {$\beta$}
\writefig   2.0 0.6 {\footnotesize $x_1 = \omega_1(0)$}
\writefig   5.9 1.05    {\footnotesize $x_2 = \omega_2(0)$}
\writefig   3.5 4.0 {\footnotesize $\omega_1(\beta)$}
}
\caption{The Feynman-Kac representation of the partition function for a gas of bosons. The
horizontal plane represents the $d$ spatial dimensions, and the vertical axis is the
imaginary time dimension.
The picture shows a situation with five particles and two cycles, with winding
numbers 4 and 1.}
\label{figfeykac}
\end{figure}

The density of particles in cycles of length $n$ is given by the expectation of the
``observable'' $\frac1V \sum_{j=1}^k n \delta_{n_j,n}$. Some computations yield the following
finite volume expression \cite{Uel}:
\be
\label{densitycyclesgdcan}
\varrho(n) = \e{\beta\mu n} \int\dd W_{00}^{n\beta}(\omega) \e{-\beta U(\omega)}
\frac{Z(\omega)}Z.
\ee
Here, $Z(\omega)$ is the grand-canonical partition function for a system where the
trajectory $\omega$ is present. It is given by the same formula as Eq.\ \eqref{FKpartfct}, but
with the additional term
$$
\prod_{j=1}^k \e{-\beta U(\omega,\omega_j)}.
$$
For repulsive interactions we always have $Z(\omega) \leq Z$, with equality in the ideal
gas. When the volume is finite we have
$$
\sum_{n\geq1} \varrho(n) = \Bigl\langle \frac NV \Bigr\rangle = \rho,
$$
where $\rho$ is the density of the system, which depends on the chemical potential. Let
$\bsvarrho(n)$ denote the thermodynamic limit of $\varrho(n)$. Limits and infinite sums do
not always commute because of a possible ``leak to infinity''. Fatou's lemma garantees
that
$$
\sum_{n\geq1} \bsvarrho(n) \leq \rho.
$$
This suggests to define the density of particles in infinite cycles by
\be
\label{defrhoinfinity}
\bsvarrho(\infty) = \rho - \sum_{n\geq1} \bsvarrho(n).
\ee

The off-diagonal correlation function of Penrose and Onsager also has an expression in the
Feynman-Kac representation. It involves an open trajectory that starts at $x$ and ends at
$y$, possibly winding several times around the time direction. Precisely, we introduce
\be
\label{odlro}
\sigma(x,y) = \sum_{n\geq1} \e{\beta\mu n} \int \dd W_{xy}^{n\beta}(\omega)
\e{-\beta U(\omega)} \frac{Z(\omega)}Z.
\ee
True to our convention we denote the thermodynamic limit by $\bssigma(x,y)$.

We immediately obtain a finite volume relation between $\varrho(n)$
and $\sigma(x,y)$. It is enough to consider $\sigma(0,x)$ because of
translation invariance (we use periodic boundary conditions). From
\eqref{densitycyclesgdcan} and \eqref{odlro}, we have \be
\label{coefficients} \sigma(0,x) = \sum_{n\geq1} c_n(x) \,
\varrho(n), \ee where the coefficients $c_n(x)$ are given by \be
\label{defcoeff} c_n(x) = \frac{\displaystyle \int\dd
W_{0x}^{n\beta}(\omega) \e{-\beta U(\omega)}
\frac{Z(\omega)}Z}{\displaystyle \int\dd W_{00}^{n\beta}(\omega)
\e{-\beta U(\omega)} \frac{Z(\omega)}Z}. \ee We need to understand
the thermodynamic limit of \eqref{coefficients}. ``Leaks to
infinity'' occur, and we cannot simply replace finite volume terms
by infinite volume ones. The ideal Bose gas yields exact expressions
and therefore constitutes a useful first step. As $V\to\infty$, Eq.\
\eqref{coefficients} becomes \cite{Uel} \be \bssigma(0,x) =
\sum_{n\geq1} \e{-x^2/4n\beta} \bsvarrho(n) + \bsvarrho(\infty). \ee

\subsection{Weakly interacting systems}

We seek to understand the general behavior of the coefficients $\bsc_n$ and
$\bsc_\infty$ in a gas of bosons with weak
interactions. We can represent the numerator and denominator of \eqref{defcoeff}
as the connectivity of a weakly self-avoiding random walk in a random environment. The
random environment is due to $Z(\omega)$, which involves sums and integrals over
configurations $\underline\omega$ of closed trajectories. We obtain
$$
\int\dd W^{n\beta}_{0x}(\omega) \e{-\beta U(\omega)} \frac{Z(\omega)}Z = \int\dd\nu(\underline\omega) \, C^{\underline\omega}_n(x),
$$
with $\nu$ a probability measure ($\int\dd\nu(\underline\omega) = 1$), and $C^{\underline\omega}_n(x)$ the connectivity of the random walk between positions 0 and $x$:
$$
C^{\underline\omega}_n(x) = \int\dd W^{n\beta}_{0x}(\omega) \e{-\beta U(\omega)} \e{-\beta U(\omega,\underline\omega)}.
$$
The latter term represents the interactions between $\omega$ and the
configuration of trajectories $\underline\omega$. The environment
typically consists of trajectories spread all over the domain, with
no significant fluctuation of density. When interactions are weak,
the typical trajectory from 0 to $x$ does not depend much on the
presence of the random environment. For $n$ large we have
$U(\omega,\underline\omega) \approx un$ for almost all $\omega$ and
$\underline\omega$, where $u$ is a positive number. It is known that
(weakly) self-avoiding random walks satisfy a local central limit
theorem in high dimensions \cite{HHS,BR}, so that, as $n\to\infty$,
\be \label{interfaibles}
\frac{C^{\underline\omega}_n(x)}{C^{\underline\omega}_n(0)} \approx
\exp \Bigl( -\frac{x^2}{4an\beta} \Bigr). \ee The constant $a$ is
greater than 1 and it reflects the (weak) self-avoidance of
$\omega$. The situation is then similar to the ideal gas, and we
expect the formula \eqref{LaFormule} to hold with $\bsc_n(x)$ given
by \eqref{interfaibles}, and $\bsc_\infty(x) = 1$ for all $x$.

\subsection{Feynman cycles in crystals}

Bosons with strong interactions are likely to undergo a usual
condensation into a solid at low temperature. We consider a
crystalline phase where particles display (diagonal) long-range
order, and where there is no off-diagonal long-range order. (We do
not consider a supersolid phase here.) We are going to argue that
infinite cycles are present at very low temperature, in dimension
three or more.
\bfig
\epsfxsize=70mm
\centerline{\epsffile{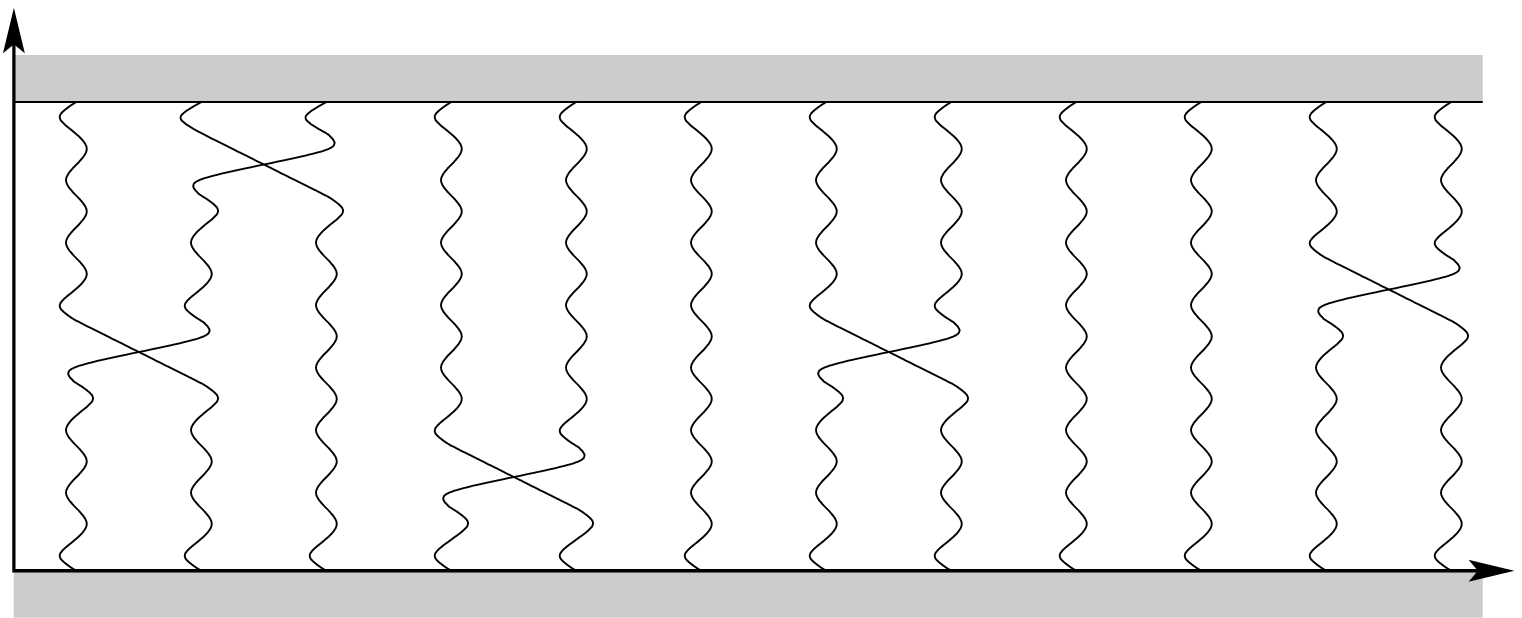}} \figtext{ \writefig   0.4
2.8 {$\beta$} } \caption{Schematic typical space-time configuration
for a crystalline phase. Trajectories are roughly vertical, with
occasional tunneling to a neighbor.} \label{figsolidfeykac}
\end{figure}
The typical Feynman-Kac representation for a quantum system in a
crystalline phase is depicted in Fig.\ \ref{figsolidfeykac}.
Particles are represented by rather straight trajectories that are given by Brownian motions in the
presence of an effective harmonic trap, due to the
neighboring particles. Occasionally, two neighboring particles may
exchange their positions through quantum tunneling.
Thus Feynman cycles are associated with the following simple
dynamical system. Particles occupy the sites of a lattice, and to a each bond is
associated a Poisson process that exchanges the corresponding particles. The rate is
proportional to $\beta$ --- the actual constant depends on
the potential barrier for switching two particles, and on the
entropy of Brownian paths that do it. Any site belongs to a Feynman cycle of arbitrary
length.

The cycle that contains a given site can also be viewed as a
continuous-time discrete random walk. From time 0 to $\beta$, jumps
on neighboring sites occur with a rate that is essentially
independent of $\beta$. If the position at time $\beta$ is identical
to the original one, the cycle closes and has winding number 1. The
position is otherwise again given by a random walk, but with the
following two restrictions: (a) No jump can occur if the resulting
space-time picture has more than one particle at a given site; (b)
if the random walk meets the other end of a bond, it has to jump in
the opposite direction. (This process appears in the study
\cite{Toth} of the Heisenberg ferromagnet.) It is illustrated in
Fig.\ \ref{figsolidcycles}. The first restriction is a
self-avoidance condition; the second restriction amounts to an
effective attraction. These should not be too important in
dimensions higher or equal to three, where random walks are
transient.

\bfig \epsfxsize=70mm \centerline{\epsffile{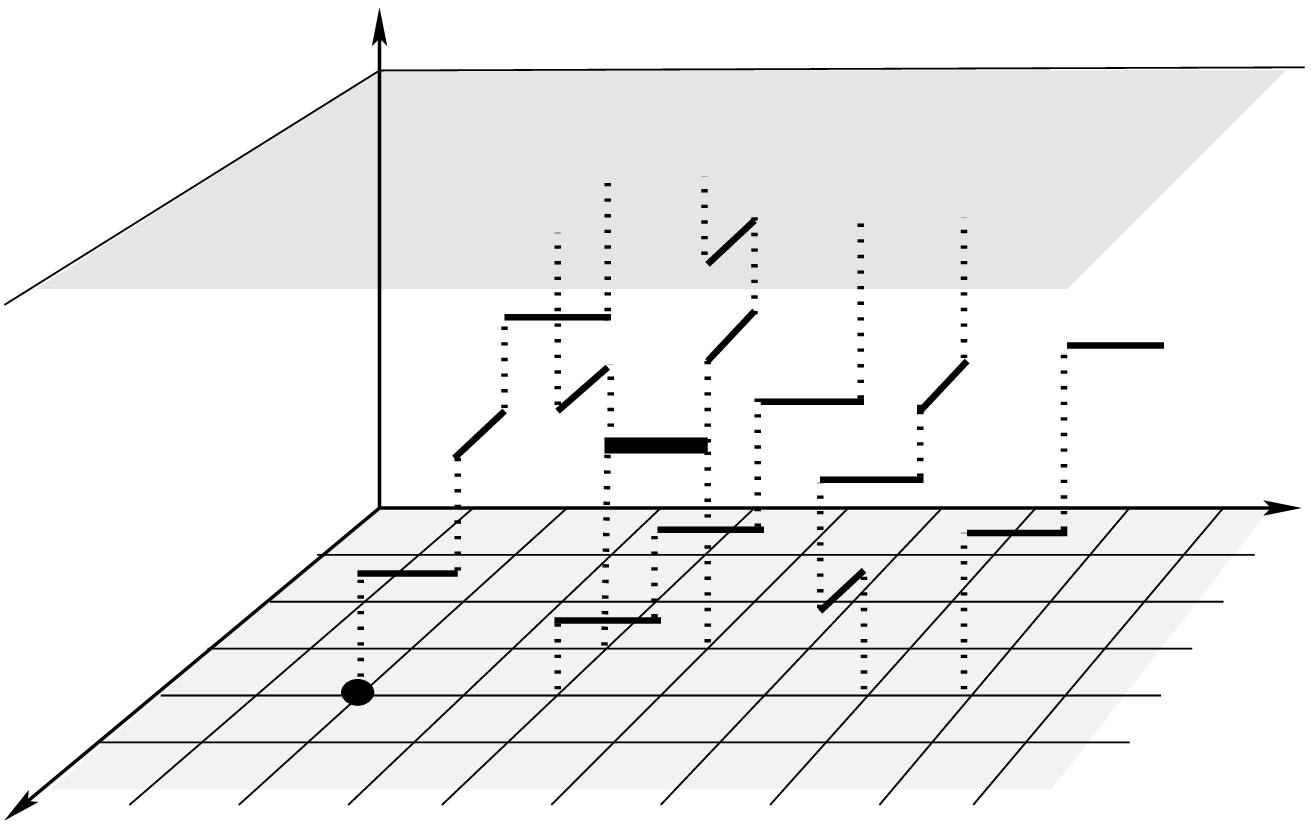}}
\figtext{ \writefig   0.45 0.35    {$x$} \writefig   7.8 2.0 {$y$}
\writefig   2.4    4.35 {$\beta$} } \caption{The Feynman cycle in a
solid resembles a discrete random walk. The thick bond is crossed in
both directions.} \label{figsolidcycles}
\end{figure}

This suggests that the typical Feynman trajectories for bosons in a
crystal (in 3D) is infinite if $\beta$ is large enough, and $\lim_{\beta\to\infty} \bsvarrho(\infty) = \rho$. On the other hand, the off-diagonal correlation function shows exponential decay in a crystal, and $\bsrho_0=0$.

Finally, let us understand the behavior of the coefficients $\bsc_n(x), \bsc_\infty(x)$ in
a crystalline phase, so as to convince ourselves that Eq.\ \eqref{LaFormule} and the
subsequent properties are still valid. For simplicity, we assume a cubic structure, but
the argument is more general. We investigate the typical space-time configurations, which
contribute the most to the numerator and denominator of Eq.\ \eqref{defcoeff}. It helps to
understand first the situation without tunneling, and then to take it into account. The
denominator typically involves crystalline space-time configurations as in Fig.\
\ref{figsolidfeykac}. The tunneling must be such that the particle located at the origin
belongs to a trajectory with winding number $n$.

\bfig \epsfxsize=70mm \centerline{\epsffile{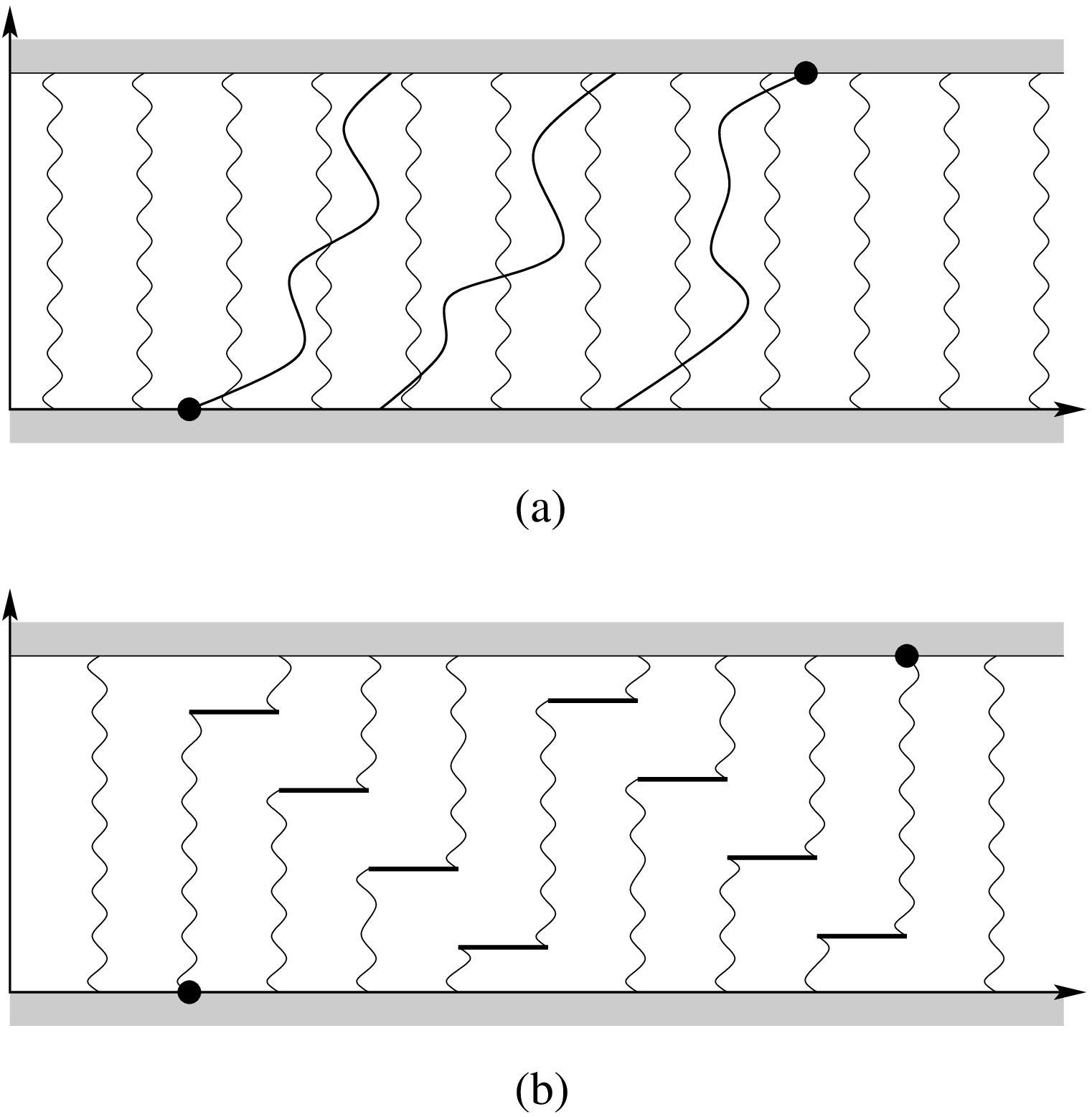}}
\figtext{ \writefig   0.4 7.0 {$\beta$} \writefig   1.84    5.13 {0}
\writefig   5.8 6.75    {$x$} \writefig   0.4 3.25    {$\beta$}
\writefig   1.83    0.8 {0} \writefig   6.45    3.5 {$x$} }
\caption{The two possibilities for the typical open trajectory from
0 to $x$. For clarity, occasional tunneling between vertical
trajectories is not shown, unlike in Fig.\ \ref{figsolidfeykac}. (a)
An extra trajectory is superimposed on the crystalline structure.
(b) The open trajectory is part of the crystalline structure.}
\label{figsolidodlro}
\end{figure}

The numerator of \eqref{defcoeff} is more complicated. One can identify two alternative typical behavior, that are depicted in Fig.\ \ref{figsolidodlro} (a) and (b). Both alternatives yield the same conclusion, namely that $\bsc_n(x) \approx \e{-a|x|}$ provided $n$ is large (or infinite). This result should be contrasted with the weakly interacting case, where $\bsc_n(x) \approx 1$ if $n \gg x^2$.

The first alternative involves an open cycle that is superimposed to the background crystalline configuration, which is left essentially undisturbed; see Fig.\ \ref{figsolidodlro} (a). Let $m$ be the winding number in absence of tunneling. The contribution of all trajectories is approximately
\be
\label{approxweight}
(4\pi\beta m)^{-d/2} \e{\beta\mu^* m} \e{-x^2/4m\beta}.
\ee
Here, $\mu^* = \mu - \langle U(\omega,\underline\omega) \rangle$ is the effective chemical potential. It is negative since the crystalline phase is stable. The maximal contribution comes from $m \approx \frac{|x|}{2\beta\sqrt{-\mu^*}}$, and the dominant term in \eqref{approxweight} is
$\e{-\sqrt{-\mu^*} |x|}$. For $n \gg |x|$ one can argue that tunneling contributes equally to
the numerator and the denominator of \eqref{defcoeff}, and therefore $\bsc_n(x) \approx \e{-\sqrt{-\mu^*} |x|}$.

The second alternative is depicted in Fig.\ \ref{figsolidodlro} (b). The open trajectory
is part of the crystalline structure. Since the distance between particles is
$\rho^{1/d}$, the open trajectory involves $|x|/\rho^{1/d}$ ``one-directional
tunnelings''. Each contributes a small factor to the ratio in \eqref{defcoeff}. Since tunneling affects equally the numerator and the denominator, we again find that $\bsc_n(x)$ decays exponentially with $|x|$, uniformly in large $n$.

As a consequence, Eq.\ \eqref{LaFormule} predicts that $\bssigma(0,x) \to 0$
as $|x|\to\infty$, while $\bsvarrho(\infty) \neq 0$.

\subsection{Conclusion}

We have proposed an exact relation, Eq.\ \eqref{LaFormule}, between Feynman cycles and
the off-diagonal correlation function. This relation involves the coefficients $\bsc_n$
which can be computed in the case of the ideal gas. It is also backed by a few rigorous results
valid for interacting systems. Our study suggests that both order parameters are identical
in the weakly interacting Bose gas, but not in strongly interacting systems in a crystalline
phase. The density of particles in infinite cycles, $\bsvarrho(\infty)$, may be strictly
positive while $\bsvarrho_0=0$. An interesting open question is whether
$\bsvarrho(\infty) > \bsvarrho_0 > 0$ for a certain choice of parameters.

I am grateful to the referees, whose comments allowed me to improve the clarity of
this letter.

\end{document}